\documentclass[12pt,graphicx]{article}
\usepackage{hyperref}

\usepackage{graphicx}
\usepackage{amssymb}

\def\vep{\varepsilon}

\def\be{\begin{equation}}
\def\ee{\end{equation}}
\def\bea{\begin{eqnarray}}
\def\eea{\end{eqnarray}}

\def\IR{\relax{\rm I\kern-.18em R}}
\def\binomial#1#2{\left(\,{\buildrel 
{\raise4pt\hbox{$\displaystyle{#1}$}}\over
{\raise-6pt\hbox{$\displaystyle{#2}$}}}\,\right)}

\def\[{\lfloor{\hskip 0.35pt}\!\!\!\lceil}
\def\]{\rfloor{\hskip 0.35pt}\!\!\!\rceil}

\def\a{\alpha}
\def\b{\beta}
\def\c{\chi}
\def\d{\delta}
\def\e{\epsilon}

\def\g{\gamma}
\def\h{\eta}

\def\j{\psi}

\def\l{\lambda}

\def\q{\theta}

\def\s{\sigma}

\def\F{\Phi}

\def\J{\Psi}
\def\L{\Lambda}
\def\O{\Omega}

\def\S{\Sigma}

\def\ca{{\cal A}}
\def\cb{{\cal B}}

\def\cg{{\cal G}}

\newcommand{\drawsquare}[2]{\hbox{%
\rule{#2pt}{#1pt}\hskip-#2pt
\rule{#1pt}{#2pt}\hskip-#1pt
\rule[#1pt]{#1pt}{#2pt}}\rule[#1pt]{#2pt}{#2pt}\hskip-#2pt
\rule{#2pt}{#1pt}}

\renewcommand{\Box}{\,\raisebox{-.45pt}{\drawsquare{7}{0.6}}\,}

\def\bo{{\raise.15ex\hbox{\large$\Box$}}}               
\def\pa{\partial}                                       
\def\TH{{\raise.2ex\hbox{$\displaystyle \bigodot$}\mskip-4.7mu \llap H
\;}}
\def\face{{\raise.2ex\hbox{$\displaystyle \bigodot$}\mskip-2.2mu \llap
{$\ddot
        \smile$}}}                                      

\def\Hat#1{\widehat{#1}}                        
\def\leftrightarrowfill{$\mathsurround=0pt \mathord\leftarrow \mkern-6mu
        \cleaders\hbox{$\mkern-2mu \mathord- \mkern-2mu$}\hfill
        \mkern-6mu \mathord\rightarrow$}
\def\dvec#1{\vbox{\ialign{##\crcr
        \leftrightarrowfill\crcr\noalign{\kern-1pt\nointerlineskip}
        $\hfil\displaystyle{#1}\hfil$\crcr}}}           

\catcode`@=11
\def\un#1{\relax\ifmmode\@@underline#1\else
        $\@@underline{\hbox{#1}}$\relax\fi}
\catcode`@=12

\def\fracm#1#2{\hbox{\large{${\frac{{#1}}{{#2}}}$}}}
\def\frac#1#2{{\textstyle{#1\over\vphantom2\smash{\raise.20ex
        \hbox{$\scriptstyle{#2}$}}}}}                   
\def\sfrac#1#2{{\vphantom1\smash{\lower.5ex\hbox{\small$#1$}}\over
        \vphantom1\smash{\raise.4ex\hbox{\small$#2$}}}} 
\def\bfrac#1#2{{\vphantom1\smash{\lower.5ex\hbox{$#1$}}\over
        \vphantom1\smash{\raise.3ex\hbox{$#2$}}}}       
\def\afrac#1#2{{\vphantom1\smash{\lower.5ex\hbox{$#1$}}\over#2}}    

\newskip\humongous \humongous=0pt plus 1000pt minus 1000pt
\def\caja{\mathsurround=0pt}
\def\eqalign#1{\,\vcenter{\openup2\jot \caja
        \ialign{\strut \hfil$\displaystyle{##}$&$
        \displaystyle{{}##}$\hfil\crcr#1\crcr}}\,}
\newif\ifdtup

  \def\pp{{\mathchoice
          {
              \kern 1pt%
              \raise 1pt
              \vbox{\hrule width5pt height0.4pt depth0pt
                    \kern -2pt
                    \hbox{\kern 2.3pt
                          \vrule width0.4pt height6pt depth0pt
                          }
                    \kern -2pt
                    \hrule width5pt height0.4pt depth0pt}%
                    \kern 1pt
           }
            {
              \kern 1pt%
              \raise 1pt
              \vbox{\hrule width4.3pt height0.4pt depth0pt
                    \kern -1.8pt
                    \hbox{\kern 1.95pt
                          \vrule width0.4pt height5.4pt depth0pt
                          }
                    \kern -1.8pt
                    \hrule width4.3pt height0.4pt depth0pt}%
                    \kern 1pt
            }
            {
              \kern 0.5pt%
              \raise 1pt
              \vbox{\hrule width4.0pt height0.3pt depth0pt
                    \kern -1.9pt  
                    \hbox{\kern 1.85pt
                          \vrule width0.3pt height5.7pt depth0pt
                          }
                    \kern -1.9pt
                    \hrule width4.0pt height0.3pt depth0pt}%
                    \kern 0.5pt
            }
            {
              \kern 0.5pt%
              \raise 1pt
              \vbox{\hrule width3.6pt height0.3pt depth0pt
                    \kern -1.5pt
                    \hbox{\kern 1.65pt
                          \vrule width0.3pt height4.5pt depth0pt
                          }
                    \kern -1.5pt
                    \hrule width3.6pt height0.3pt depth0pt}%
                    \kern 0.5pt
            }
        }}

  \def\mm{{\mathchoice
                       {
                             \kern 1pt
               \raise 1pt    \vbox{\hrule width5pt height0.4pt depth0pt
                                  \kern 2pt
                                  \hrule width5pt height0.4pt depth0pt}
                             \kern 1pt}
                       {
                            \kern 1pt
               \raise 1pt \vbox{\hrule width4.3pt height0.4pt depth0pt
                                  \kern 1.8pt
                                  \hrule width4.3pt height0.4pt depth0pt}
                             \kern 1pt}
                       {
                            \kern 0.5pt
               \raise 1pt
                            \vbox{\hrule width4.0pt height0.3pt depth0pt
                                  \kern 1.9pt
                                  \hrule width4.0pt height0.3pt depth0pt}
                            \kern 1pt}
                       {
                           \kern 0.5pt
             \raise 1pt  \vbox{\hrule width3.6pt height0.3pt depth0pt
                                  \kern 1.5pt
                                  \hrule width3.6pt height0.3pt depth0pt}
                           \kern 0.5pt}
                       }}

\def\dslash{\not{\hbox{\kern-2pt $\partial$}}}
\def\Dslash{\not{\hbox{\kern-4pt $D$}}}
\def\pslash{\not{\hbox{\kern-2.3pt $p$}}}
 \newtoks\slashfraction
 \slashfraction={.13}
 \def\slash#1{\setbox0\hbox{$ #1 $}
 \setbox0\hbox to \the\slashfraction\wd0{\hss \box0}/\box0 }
 
\font\ro=cmsy10                          
\def\kcr{{\hbox{\ro \char'170}}}                
\def\ktl{{\hbox{\ro \char'170}}}        
\def\ktr{{\hbox{\ro \char'170}}}        
\def\kbl{{\hbox{\ro \char'170}}}        
\def\kbr{{\hbox{\ro \char'170}}}        

\def\plpl{\raise-2pt\hbox{$\raise3pt\hbox{$_+$}\hskip-6.67pt\raise0.0pt
\hbox{$^+$}\hskip 0.01pt$}}
\def\mimi{\raise-2pt\hbox{$\raise3pt\hbox{$_-$}\hskip-6.67pt\raise0.0pt
\hbox{$^-$}\hskip 0.01pt$}}

\parskip=\medskipamount                 
\thicklines                         

\def\border{                                            
        \setlength{\unitlength}{1mm}
        \newcount\xco
        \newcount\yco
        \xco=-21
        \yco=12
        \begin{picture}(140,0)
        \put(\xco,\yco){$\ktl$}
        \advance\yco by-1
        {\loop
        \put(\xco,\yco){$\kcr$}
        \advance\yco by-2
        \ifnum\yco>-240
        \repeat
        \put(\xco,\yco){$\kbl$}}
        \xco=158
        \yco=12
        \put(\xco,\yco){$\ktr$}
        \advance\yco by-1
        {\loop
        \put(\xco,\yco){$\kcr$}
        \advance\yco by-2
        \ifnum\yco>-240
        \repeat
        \put(\xco,\yco){$\kbr$}}
        \put(-20,13){\tiny **University of Maryland * Center for String and 
         Particle  Theory* Physics Department***University of Maryland *Center  
        for String and Particle  Theory** }
        \put(-20,-241.5){\tiny **University of Maryland * Center for String and 
         Particle  Theory* Physics Department***University of Maryland *Center  
        for String and Particle  Theory** }
        \end{picture}
        \par\vskip-8mm}
\def\headpic{                                           
        \indent
        \setlength{\unitlength}{.4mm}
        \thinlines
        \par
        \begin{picture}(29,16)
        \put(165,16){\line(1,0){4}}
        \put(170,16){\line(1,0){4}}
        \put(180,16){\line(1,0){4}}
        \put(175,0){\line(1,0){4}}
        \put(180,0){\line(1,0){4}}
        \put(185,0){\line(1,0){4}}
        \put(169,0){\line(0,1){16}}
        \put(170,0){\line(0,1){16}}
        \put(179,0){\line(0,1){16}}
        \put(180,0){\line(0,1){16}}
        \put(184,0){\line(0,1){16}}
        \put(185,0){\line(0,1){16}}
        \put(169,16){\oval(8,32)[bl]}
        \put(170,16){\oval(8,32)[br]}
        \put(179,0){\oval(8,32)[tl]}
        \put(185,0){\oval(8,32)[tr]}
        \end{picture}
        \par\vskip-6.5mm
        \thicklines}
\def\title#1#2#3#4{\border\headpic {\hbox to\hsize{#4 \hfill UMDEPP #3}}\par
        \begin{center} \vglue .5in {\large\bf #1}\\[.6in]
        {#2}\\[.1in] {\it Department of Physics and Astronomy}\\
        {\it University of Maryland, College Park, MD 20742}\\[1.5in]
        {\bf ABSTRACT}\\[.1in] \end{center} \begin{quotation}}  
\def\Title#1#2#3#4#5#6#7{\border\headpic
        {\hbox to\hsize{#7 \hfill UMDEPP #6}}\par
        \begin{center} \vglue .4in {\large\bf #1}\\[.4in]
        {#2}\\[.1in] {\it Department of Physics and Astronomy}\\
        {\it University of Maryland, College Park, MD 20742}\\[.1in]
        {#3}\\[.1in] {\it {#4}}\\ {\it {#5}}\\[.5in] {\bf ABSTRACT}\\[.1in]
        \end{center} \begin{quotation}}                 
\def\endtitle{\end{quotation}\newpage}                  

\def\qd{{\kern0.5pt
                   q \kern-5.05pt \raise5.8pt\hbox{$\textstyle.$}\kern
0.5pt}}

\makeatletter
    \@addtoreset{equation}{section}%
\makeatother

\begin{document}
\def\gfrac#1#2{\frac {\scriptstyle{#1}}
        {\mbox{\raisebox{-.6ex}{$\scriptstyle{#2}$}}}}
\def\gg{{\hbox{\sc g}}}
\border\headpic 
{\hbox to\hsize{November 2003 \hfill UMDEPP 04-009}}
{\hbox to\hsize{~\hfill UFIFT-HEP 03-31}}
\par
\setlength{\oddsidemargin}{0.3in}
\setlength{\evensidemargin}{-0.3in}
\begin{center}
\vglue .10in
{ \large\bf 
Field Strengths of Linearized    }
\\[.05in]
{ \large\bf 
5D, $\cal N$ = 1 Superfield Supergravity  }
\\[.05in]
{ \large\bf 
On a 3-Brane\footnote
{Supported in part  by National Science Foundation Grant
PHY-0099544.}\  }
\\[.5in]
S. James Gates, Jr.$^\dag$\footnote{gatess@wam.umd.edu}, William D. Linch,
III$^\dag$\footnote{ldw@physics.umd.edu} and J.
Phillips$^{\dag\star}$\footnote{ferrigno@physics.umd.edu}
\\[0.06in]
{\it ${}^\dag$Center for String and Particle Theory\\ 
Department of Physics, University of Maryland\\ 
College Park, MD 20742-4111 USA}\\
~\\
{\it and}\\
~\\
{\it ${}^\star$Institute for Fundamental Theory\\
Department of Physics, University of Florida,\\
 Gainesville, FL, 32611, USA}\\
[.5in]

{\bf ABSTRACT}\\[.01in]
\end{center}
\begin{quotation}
{Recently a description of linearized 5D supergravity in 4D, ${\cal N} =1$ superspace was presented.  By analyzing the on-shell component Lagrangian, this description was proven to be Lorentz invariant in five dimensions.  This paper describes a geometric formulation of the 5D supergravity in 4D, ${\cal N} =1$ superspace. Using this new formalism, the on-shell structure of the previous description is verified from purely superspace methods.  This geometric description provides a connection to dimensionally reduced manifestly supersymmetric 5D supergravity.}

${~~~}$ \newline
\endtitle

\section{Introduction}

$~~~\,~$About a decade ago, Marcus et. al. \cite{Marcus:1983wb} advocated the utility of 
writing higher dimensional supersymmetric theories in the language of 4D, ${\cal N} = 1$ (a.k.a. ``simple") superspace.  They presented 10D super-Yang-Mills theory and studied various aspects of its quantum behavior. This approach was more recently rediscovered by Arkani-Hamed et. al. \cite{Arkani-Hamed:2001tb} who studied super-Yang-Mills theory in various dimensions $D>4$, for the purpose of studying the phenomenology of certain quantum effects in brane world scenarios.  

More recently,  a model of linearized 5D supergravity \cite{Linch:2002wg} was presented within the framework of simple superspace.  The bosonic part of this theory was shown to be the bosonic part of 5D component supergravity on-shell and represents a minimal extension of linearized old minimal ($n=-{\frac 13}$) supergravity.  This model was subsequently used to study gravity mediated supersymmetry breaking in the simplest 5D brane world scenario \cite{Buchbinder:2003qu}.  The same scenario was investigated by other authors in the component formalism \cite{Rattazzi:2003rj} and exactly the same coefficients were obtained for the one-loop operators in the effective action, yielding another check of the validity of the 4D, ${\cal N}=1$ superspace formalism.

Theories of this nature are interesting for various reasons. One such motivation is to clarify the mechanism of dimensional reduction in superspace. While dimensionally reducing component results is a simple and powerful procedure, the analogue in manifestly supersymmetric theories is not understood nearly as well.  A second reason for studying massless theories in 5D dimensions is that they are related to massive higher spin theories in four dimensions.   Lagrangian formulations of supersymmetric massive higher spin theories in 4D have only recently been investigated \cite{Massive1,Massive2,Massive3}, and a formulation for all higher spins still remains an open question. It is, furthermore, common to use lower ${\cal N}$ superfields to model higher ${\cal N}$ supersymmetry in supersymmetric quantum field theory.  For the most recent applications to quantum higher ${\cal N}$ models see \cite{Quantum1,Quantum2,Quantum3,Quantum4,Quantum5}.

While the action principle given in \cite{Linch:2002wg} suffices to study the dynamics of the theory, there are various unsatisfactory aspects of this presentation.  An analysis of the component projection was made to show that, after one integrates out the full set of auxiliary fields, the bosonic sector of the theory describes linearized gravity coupled minimally to Maxwell theory.  Although component projection is straightforward, it is cumbersome.  Additionally, in Wess-Zumino gauge, the theory contains extra fields that have the correct mass dimension to propagate.  Fortunately, these extraneous fields are not present in the on-shell Lagrangian. Furthermore, the approach of \cite{Linch:2002wg} sheds little light on the problems of possible non-linear extensions of the 5D theory, higher dimensional analogues, or curved backgrounds such as AdS${}_5$.  

The aim of this report is to clarify the geometric structure of the theory presented in \cite{Linch:2002wg}.  As we will demonstrate, this exposition has the advantage that all claims about the on-shell structure of the theory can be verified using solely superspace technology, without recourse to component calculations.  In particular, the unknown coefficient of \cite{Linch:2002wg} and the on-shell analysis can be obtained without resorting to component arguments.   We further hope that having a geometric formulation at our disposal will expedite the development of the less trivial non-linear and higher dimensional theories.  In the future, this geometric description may be smoothly mapped into a dimensionally reduced geometry coming from a manifestly supersymmetric 5D geometry.  The method used to develop our geometric prescription was inspired by \cite{Gates:1979gv}.

We organize this paper as follows.  The discussion is presented in two sections.  In the first section, we present the 5D Maxwell limit of \cite{Marcus:1983wb} where the geometric description is easy to follow.  We then proceed with the analysis of 5D linearized supergravity.  Since the method developed in this section generates a minimal set of field strengths, the second section is devoted to building a full geometric description.  When describing manifestly 5D Lorentz covariant objects we use capital calligraphic letters, $\cal A,B...$ to represent vector indices and tilde Greek letters, $\tilde\a, \tilde\b...$, to represent spinor indices.  For all dimensionally reduced 4D objects we use the notation and conventions of \cite{Buchbinder:qv}, along with the identification $\un a=\a\dot\a$.  

\section{Superspace Geometry and Dynamics}
$~~~\,~$The outline of this section is the following.  We start with an embedding of the gauge fields into ${\cal N} = 1$ superfields.  With this information, we construct a set of field strengths and work out a set of identities relating these field strengths.  The next step involves writing the equations of motion in terms of the field strengths.  Finally, we determine the propagating field strengths and show that, on-shell, these describe the correct propagating degrees of freedom.
\subsection{Super Maxwell Theory}
$~~~\,~$We start with 5D super-Maxwell theory.  The minimal 5D super-Maxwell theory is a theory of three dynamical component field strengths: the  vector field strength $F_{\cal {AB}}$, a spinor field strength $\l_{\tilde \a}$ and an additional scalar field strength $\varphi$. (A simple way to see the need for this scalar is to count the physical degrees of freedom. $\l_{\tilde \a}$ has 4 physical degrees of freedom and the 5D gauge vector has 3.  Hence, supersymmetry demands one more bosonic degree of freedom.) In terms of 4D variables, the vector field strength splits into a 4D vector field strength $F_{\un a \, \un b}$ and a 5-component $F_{{\un a} \, 5}$. The spinor field $\l_{\tilde \a}$ splits into two spinor field strengths $\l^{(\pm)}_\a$  where the ($\pm$) here refers to $x^5$ parity.  We define this operation by acting on 5D spinors 
with projection operators $P_{\pm}$ (thus breaking 5D Lorentz symmetry) $P_{\pm} 
:= \fracm 12 ( I\pm i \g^5  )$.

Since ${\cal N} = 1$ superspace produces results in 4D spin-tensor notation, we need to write the Bianchi identity and equation of motion for the vector field strength in this notation. This analysis is useful for both Maxwell theory and supergravity, since both have a gauge vector in their component spectra. The 4D vector field strength has the usual decomposition:
\bea
F_{\un a \, \un b} ~=~ \e_{\dot\a\dot\b} f_{\a\b} ~+~ \e_{\a\b}\bar f_{\dot\a\dot\b}~~~,
\eea
so the equation of motion $\pa^{\cal A}  F_{{\cal A} \, {\cal B}}=0$ becomes:
\bea
\label{nosusyeom1}
\pa^{\cal A}F_{{\cal A} \, \un b} ~=~ \pa_5F_{5\un b} ~+~ \frac 12 \, \pa_\b{}^{\dot\a}\bar f_{\dot\a\dot\b}
~+~ \frac 12 \, \pa_{\dot\b}{}^\a\ f_{\a\b} ~=~ 0~~~,\\
\label{nosusyeom2}
\pa^{\ca}F_{\ca5}~=~ \frac 12 \, \pa^{\un a}F_{5\un a} ~=~ 0 ~~~,
\eea
and the Bianchi identity, $\pa_{[A}F_{BC]}=0$, takes the form
\bea
\label{nosusybi1}
\pa_{[5}F_{\un b\un c]} ~=~ \e_{\dot\b\dot\g}(2\pa_5f_{\b\g}
~-~ \pa_{(\g}{}^{\dot\g}F_{5\b)\dot\g}) ~+~ {\rm c.c.} ~=~ 0~~~,\\
\label{nosusybi2}
\pa_{[\un a}F_{\un b\un c]} ~=~ \e_{\b\g}\e_{\dot\a(\dot\b}\d_{\dot\g)}{}^{\dot\s}
(\pa_\a{}^{\dot\d}\bar f_{\dot\s\dot\d}-\pa_{\dot\s}{}^\d f_{\d\a})
~+~ {\rm c.c.}=0~~~.
\eea

The simple superspace embedding of the gauge potentials goes as follows.  The 4D piece of the vector potential and the $(+)$ piece of the 5D spinor are embedded in a real scalar superfield, $V$.  The $(-)$ piece of the 5D spinor, the fifth
component of the vector, and the scalar are all contained in a chiral scalar field, $\Phi$.  The gauge transformations are:
\bea
\label{Maxgauge}
\d V=-\frac i2(\L-\bar \L )\hspace{1cm},\hspace{1cm} \d\F=\pa_5 \L\hspace{1cm},\hspace{1cm} \bar D_{\dot \a} \L=0~~~.
\eea
In a Wess-Zumino gauge, it is easy to see that these gauge transformations contain the correct component transformation laws.  There are two obvious field strengths:
\bea
\label{Maxfs}
{ W}_\a :=-\frac14\bar D^2 D_\a V \hspace{1cm},\hspace{1cm} 
{ F}:=-\frac i2 (\F-\bar \F)-\pa_5 V~~~.
\eea
The first field strength is the usual 4D Maxwell field strength, while the second is the field strength of a 4D gauge super 0-form \cite{GGRS} covariantized to five dimensions.  With these definitions we have the following identities:
\bea
\label{Maxbasic1}
D^\a { W}_\a ~-~ \bar D_{\dot \a}\bar { W}^{\dot \a}~=~ 0~~~,\\
\label{Maxbasic}
\pa_5 { W}_\a~-~ {\frac 14} \bar D^2 D_\a { F}~=~ 0~~~,
\eea

Noting the mass dimension of these field strengths, we are led uniquely to the following Lagrangian (density):
\bea
\label{Maxlag}
L\, &=& \, {1\over 2}\int d^2\q  \, { W}^\a { W}_\a ~+~ {a\over 2}\int d^4\q \, { F}^2~~~.
\eea
Here $a$ is an unknown real coefficient.  Since the two terms in this Lagrangian are separately invariant, the gauge invariance does not allow us to fix this coefficient, rather, it is 5D Lorentz invariance which uniquely determines $a$. Previously, the only approach to fix $a$ was to perform the component projection of the Lagrangian (\ref{Maxlag}), integrate out the auxiliary fields and assemble the bosonic components into the 5D kinetic term $F_{\ca\cb}F^{\ca\cb}$, which can be done correctly iff $a=2$. Although the component analysis here is straightforward, it becomes increasingly awkward for more complicated theories.  One of the main points of this paper is to give a procedure for fixing this coefficient  (and similar ones encountered in more complicated theories) using superspace arguments, thereby avoiding a much more complicated component analysis.

Our procedure involves putting the field strengths on-shell by implementing the superspace equations of motion.  Once on-shell, we can show that the components of the field strengths form representations of the 5D Poincar\'e algebra for only one value of $a$.  The equations of motion are:
\bea
\label{Maxeom1}
{{\d S}\over{\d V}}&=& -D^\a { W}_\a+a \pa_5 { F}=0~~~,\\
\label{Maxeom2}
{{\d S}\over{\d \Phi}}&=&{i\over8}a\bar D^2 { F}=0~~~.
\eea
The second equation implies that ${ F}$ is a linear superfield, $D^2 { F}=0$.  To determine the coefficient $a$ we take ${\frac 18} \bar D^2D_\a$ on (\ref{Maxeom1}) and use the identity (\ref{Maxbasic}):
\be
\label{Maxrep1} \eqalign{
0~&=~ {\frac 18} \bar D^2D_\a(-D^\b { W}_\b ~+~ a \pa_5 { F}) \cr
~&=~ \Box { W}_\a  ~+~ {\frac a8}\pa_5\bar D^2 D_\a { F} \cr
~&=~ (\Box +{\frac a2}\pa_5^2){ W}_\a~~~ .
}
\ee
Here we see that for ${ W}_\a$ to be a non-trivial representation of the 5D Poincar\'e algebra, we must set $a=2$. It is only for this value that the 5D d'Alembertian
appears in the final result in (\ref{Maxrep1}).  The attentive reader may obtain a sense
of uneasiness at this result.  After all, a 5D spinor should obey a 5D Dirac equation
that is the ``square root'' of 5D d'Alembertian.  This issue is easily rendered a moot
point as we show below.

With the free parameter fixed, we proceed to prove that the components of the field strengths obey the correct 5D equations of motion.  We can see that the lowest component 
of ${ F}$ propagates appropriately to describe a 5D massless scalar by contracting $\frac 12D^\a$ on (\ref{Maxbasic}) and substituting (\ref{Maxeom1}):
\be
\label{Maxrep0} \eqalign{
0&=~ {\frac 12}D^\a \left( \pa_5 { W}_\a-{\frac 14}\bar D^2 D_\a { F}\right) \cr
&=~ {\frac 12} \pa_5 D^\a { W}_\a +\Box { F}  \cr
&=~ (\Box + \pa_5^2 ){ F}~~~.  }
\ee
So, the lowest component of ${ F}$ is the scalar degree of freedom $\varphi$.  The spinor component $D_\a { F}|$ should be $\l_\a^-$ so we look for the 5D Dirac equation.  If we take $\frac 12\bar D_{\dot\a}$ on (\ref{Maxeom1}) we get the desired result:
\be  \eqalign{
0&=~ \frac 12 \, \bar D_{\dot\a} (-D^\a { W}_\a ~+~ 2\pa_5 { F})  \cr
&=~ - i \pa_{\un a} { W}^\a+\pa_5\bar D_{\dot\a}{ F}~~~.
}   \ee
To see why this is the appropriate result, we may approach this issue from a different
viewpoint.  A massless 5D spinor $\l_\a$ must obey the 5D Dirac equation $0 ~=~ i\g^{\cal A} \, \pa_{\cal A} \, \l$ and upon multiplying by $- P_-$ we find
\be 
\label{5DDiraceom} \eqalign{
0 &=~ - \, iP_- \, \g^{\cal A} \, \pa_{\cal A} \, \l ~=~ - \, i \, P_- \, \g^{ a} \, \pa_{a} \, \l 
~-~ iP_- \, \g^{5} \, \pa_{5} \, \l   \cr
0 &=~ - \,  i \,  \g^{a} \, \pa_{a} \, P_+ \, \l 
~+~   \pa_{5} \, P_- \, \l    \cr
0 &=~ - \,  i \,  \g^{ a} \, \pa_{ a} \,  \l^{(+)} 
~+~   \pa_{5} \,  \l^{(-)}  
~~~,  }
\ee
and hence the two parts of the 5D spinor reside in two different 4D, $\cal N$ = 1 superfields.

To find the vector field strength, we note that the identity (\ref{Maxbasic}) looks similar to the Bianchi identity (\ref{nosusybi1}) after symmetrizing\footnote{We adhere to the convention $T_{(\a\b)}:=T_{\a\b}+T_{\b\a}$ which differs from the convention of \cite{Buchbinder:qv} by a normalization factor.} with $-\frac i2D_\b$:
\be
\label{SupMaxBI1}   \eqalign{
0&=~ - i \, \frac 12 \, \pa_5 D_{(\b}{ W}_{\a)} ~+~ {\frac i4} D_{(\b}\bar D^2 D_{\a)} { F} \cr
&=~ - \, i \frac 12 \, \pa_5D_{(\b}{ W}_{\a)}
~+~  {\frac 14} \,  \pa_{(\a}{}^{\dot\a}[D_{\b)},\bar D_{\dot\a}]{ F}
~~~.   }
\ee
Also, the superfield equation of motion (\ref{Maxeom1}) looks similar to the 5D Maxwell equation (\ref{nosusyeom1}) under $\frac 12[D_\a,\bar D_{\dot\a}]$:
\be
\label{SupMaxEom1}  \eqalign{
0&=~ \frac 12 \\, [D_\a,\bar D_{\dot\a}](-D^\a { W}_\a ~+~ 2 \pa_5 { F})  \cr
&=~ i \frac 12   \pa_{\dot\a}{}^\b D_{(\b}{ W}_{\a)}
~+~ i \frac 12  \,\pa_\a{}^{\dot\b}\bar D_{(\dot\b}\bar { W}_{\dot\a)}
~+~ \pa_5 [D_\a,\bar D_{\dot\a}]{ F}~~~  }
\ee
Here we have used (\ref{Maxbasic1}) twice.  From the (\ref{SupMaxBI1}) and (\ref{SupMaxEom1}), we can read off the component definitions for the vector field strength. The list of all component field strengths is
\bea
\label{Maxcomp}
f_{\a\b}=-  i \, \frac 12D_{(\b}{ W}_{\a)}|~~~~
F_{5\un a}=-\frac 12 [D_\a,\bar D_{\dot\a}]{ F}|~~~,\cr
\l_\a^{(+)}={ W}_\a|~~~~\l_\a^{(-)}=\bar D_{\dot\a}{ F}|~~~,~~~~~~\cr
\varphi ={ F}|~~~.~~~~~~~~~~~~~~~~~~~~~
\eea
Since ${ F}$ is a linear field, it has no other components.  Also, the component $D^\a { W}_\a|$ is related to ${ F}|$ by (\ref{Maxeom1}) and is, therefore, not a separate degree of freedom.  

This concludes the description of 5D Maxwell theory.  We have shown that the arbitrary coefficient $a$ can be fixed using standard superspace representation theory without resorting to component analysis.  We have also shown that the theory describes the dynamics of a vector field strength, a scalar and a spinor in five dimensions.

\subsection{Linearized Supergravity}
$~~~\,~$Armed with the preceding analysis of super-Maxell theory in five dimensions, we now follow the steps described above for the case of the minimal extension of linearized old minimal supergravity to five dimensions.  The 5D supergravity component spectrum is comprised of three dynamical field strengths:  the Weyl tensor $C_{\cal ABCD}$, the curl of the gravitino $f_{{\cal AB}\tilde\g}$, and the photon field strength $F_{\cal AB}$.  We wish to verify that the description of 5D supergravity in simple superspace given in \cite{Linch:2002wg} describes the dynamics of only these degrees of freedom.  To begin, we rewrite the field strengths in 4D spin-tensor notation.  We will carefully count the dimension of the representations of each piece of the field strengths to make sure that we do not miss anything when translating to 4D notation.

The Weyl tensor has the same symmetries as the curvature tensor.  It is also completely traceless in five dimensions:
\bea
0=\h^{\cal {A \, C}}C_{\cal {A \, B\, C\, D}}=\h^{ac}C_{ a \, {\cal B}\,  c \, {\cal D}}+\h^{55}C_{5 \, {\cal B}\, 5 \, {\cal C}}~~~.
\eea
This means that the 4D trace of $C_{a \, b \, c \, d}$ is $-C_{a \, 5 \,
c \, 5}$ and not a separate dynamical field strength.  Further, $C_{a \, b \,
c \, 5}$ and $C_{a \, 5 \, c \, 5}$ are both 4D traceless.  We have the usual result for the spinor decomposition of $C_{a \, b \,  c \, d}\,=\,{\bf 21}\,\ominus\,{\bf 10}\,\ominus\,{\bf 1}\,\cong\,
{\bf 5}\,\oplus\, \bar {\bf 5}\,=\,C_{\a\b\g\d}\,\oplus\, \bar C_{\dot \a \dot \b \dot \g \dot \d}$.  Since $C_{a \, 5 \, b \, 5}$ is symmetric and traceless we have: $C_{a \, 5 \, b \, 5}\,=\,{\bf 10}\,\ominus\,{\bf 1} \,\cong\,{\bf 3}\,\otimes\,\bar {\bf 3}\,=\, C_{\a\b\dot\a\dot\b}$.  Since $C_{a \, b \, c \, 5}$ is traceless we find: $C_{a \, b \, c \, 5}\,=\,{\bf 6}\,\otimes\,{\bf 4}\,\ominus\,{\bf 4}\,\ominus\,{\bf 4}\,\cong\,(\bar {\bf 4}\, \otimes\, {\bf 2})\,\oplus\, ({\bf 4}\,\otimes\, \bar {\bf 2}) \,=\,\bar C_{\dot \a\dot \b\dot \g \d}\,\oplus\,  C_{\a\b\g\dot\d}$.  The $\bf 10$ subtraction from $C_{a \, b \, c \, d}$ and the $\bf4$ subtraction from $C_{a \, b
\, c  \, 5}$ account for the condition $C_{[{\cal {A \, B \, C \, D}}]}=0$ inherited from the Riemann tensor in the presence of a metric.

The decomposition of the curl of the gravitino goes as follows.  First, $f_{{\cal {A\, B}} \tilde\g}$ is gamma traceless in five dimensions:
\bea
0~=~ (\g^{\cal A})_{\tilde\a}{}^{\tilde\b}  f_{{\cal {A\, B}} \tilde\b} ~=~
(\g^a)_{\tilde\a}{}^{\tilde\b}f_{a{\cal B} \tilde\b}
~+~ (\g^5)_{\tilde\a}{}^{\tilde\b}f_{5{\cal B} \tilde\b}~~~.
\eea
This means that there are $({\bf 10}\,\otimes\,{\bf 4})\,\ominus\,({\bf 5}\,\otimes\,{\bf 4})={\bf 20}$ complex degrees of freedom.  Using the basis:
\bea
(\g^a)_{\tilde\a}{}^{\tilde\b}=\left(\begin{array}{cc}
0 & (\s^a)_{\a\dot\b} \\
(\tilde\s^a)^{\dot\a\b} & 0 \\
\end{array}\right)~~~,~~~
(\g^5)_{\tilde\a}{}^{\tilde\b}=\left(\begin{array}{cc}
iI & 0 \\
0 & -iI \\
\end{array}\right)~~~,
\eea
for the 5D gamma matrices, we see that $f_{\un a \, 5 \, \b}^{(+)}$ and $f_{\un a \, 5 \, \b}^{(-)}$ 
are 4D sigma traceless.  Further, $f_{\un a \, 5 \, \b}^{(\pm)}$ is the 4D sigma trace of $f_{\un 
a \, \un b \, \dot\g}^{(\mp)}$.  This last results means that we have the usual decomposition 
for $f_{\un  a \, \un b \, \g}^{(\pm)}$ and we must decompose $f_{5 \, \un b \, \g}^{(\pm)}$ in 
terms of 4D irreducible spin tensors.  Thus, the curl of the gravitino is described by:
\bea 
f_{ab\tilde\g}\,=\,
({\bf 6}\,\otimes\,{\bf 4})\,\ominus\,({\bf5}\,\otimes\,{\bf 4})\,\cong\,{\bf 4}\,\oplus\,\bar {\bf 4}\,=\,f_{\a\b\g}^{(+)}\,\oplus\, f_{\dot\a\dot\b\dot\g}^{(-)}~~~,\cr
f_{5a\tilde\g}\,=\,({\bf4}\,\otimes\,{\bf 4})-{\bf 4}\,\cong\,({\bf 3}\,\otimes\,\bar {\bf 2})\,\oplus\, (\bar {\bf 3}\,\otimes \,{\bf 2})= f_{\a\b\dot \g}^{(-)}\,\oplus \,f_{\dot\a\dot\b\g}^{(+)}~~~.
\eea
Note that the spin tensors have a total of ${ 20}$ complex degrees of freedom.

We have just demonstrated that the $SL(2,\mathbb C)$ decomposition of the 5D on-shell supergravity multiplet is given by the following set of 4D
irreducible spin tensors
\bea
\label{cfs1}
C_{\a\b\g\d}~~,~~C_{\a\b\dot\a\dot\g}~~,~~C_{\a\b\g\dot\d}~~~,\cr
f_{\a\b\g}^{(+)}~~,~~f_{\a\b\dot\g}^{(-)}~~,~~
f_{\dot\a\dot\b\g}^{(+)}~~,~~f_{\dot\a\dot\b\dot\g}^{(-)}~~~,\cr
f_{\a\b}~~,~~F_{\un a 5}~~~.~~~~~~~~~
\eea
We now take the theory presented in \cite{Linch:2002wg}, and show that, on-shell, it describes only these field strengths.  As in the Maxwell theory, we will use the gauge transformations to construct field strengths.  Then, we present some identities that relate the various field strengths.  Finally, we will put the theory on-shell and determine the propagating degrees of freedom.  

The 5D supergravity presented in \cite{Linch:2002wg} is described by four prepotential superfields: a real vector valued superfield $H_{\un a}=\bar H_{\un a}$, a real scalar superfield $P=\bar P$, an unconstrained spinor superfield $\J_\a$, and a chiral scalar $T$.  $P$ is a potential for a chiral compensator $\S$. The latter is therefore {\it {not}} the chiral compensator
of old minimal supergravity.  Instead this form of the minimal off-shell 4D, $\cal N$ =
1 supergravity is a variant formulation of the theory wherein one of the usual spin-0
auxiliary fields is replaced by the field strength of a 3-form\footnote{This formulation
of minimal off-shell 4D, $\cal N$ = 1 supergravity was first suggested in 1980
\cite{Var}.}
: $\S=-{\frac 14}\bar D^2 P$.  The gauge transformations of these fields are:
\bea
\label{gravgauge}
\d H_{\un a}=\bar D_{\dot \a}L_\a - D_\a \bar L_{\dot \a} \hspace{0.5cm}&;&\hspace{0.5cm} \d P=D^\a L_\a+ \bar D_{\dot \a} \bar L^{\dot \a} ~~,\cr
\d \J_\a=\pa_5 L_\a -{\frac 14}D_\a \O -{\frac i4}\bar D^2 D_\a L \hspace{0.5cm}&;&\hspace{0.5cm} \d T=\pa_5 \O ~~.
\eea
Here $L_\a$ is the unconstrained gauge parameter superfield familiar from all 4D supergravity theories, $\O$ is a chiral superfield, and $L=\bar L$ is a real superfield.  (The $L$ term was discovered after the submission of \cite{Linch:2002wg}.  Although it seems to play an important role in theories in dimensions higher than five, it's existence does not affect any of the conclusions reached in the aforementioned reference.)  In a Wess-Zumino gauge, the 4D graviton and $(+)$-gravitino are contained in $H_{\un a}$.  The 4D gauge vector, metric vector component $g_{\un a \, 5}$, the $(-)$-gravitino and $(-)$-``gaugino", $\Psi_{5\a}^{(-)}$, are all components of $\J_\a$.  The scalars $A_5$ and $g_{55}$ and the $(+)$-gaugino $\Psi_{5\a}^{(+)}$ are components of $T$.  Unfortunately, there are other components which have the correct mass dimension to propagate and which can not be set to zero in Wess-Zumino gauge, viz. the vector component of $P$ and the 2-form component of $\J_\a$. Proving that these fields do not represent propagating degrees of freedom is one of the motivations for this study.

We can now construct the field strengths associated with the gauge prepotentials.  Three field strengths are inherited from old minimal supergravity in 4D:
\be
\label{gravfs1a}  \eqalign{
{ W}_{\a\b\g} &:=\, i {\frac 1{8\cdot 3!}}\bar D^2 D_{(\a} \pa_\b{}^{\dot \a} H_{\g)\dot \a} ~~~, 
\cr
\label{gravfs1b}
{ G}_{\un a} &:= ~{\frac 18}D^\b\bar D^2 D_\b H_{\un a} ~-~ {\frac 1{24}}[D_\a, \bar D_{\dot\a}][D_\b,\bar D_{\dot \b}] H^{\un b}  \cr
&~~~~~~-~{\frac 12}\pa_{\un a}\pa_{\un b}H^{\un b} ~+~ {\frac i{12}} \pa_{\un 
a}(\bar D^2-D^2)P \cr 
\label{gravfs1c}
{ R} &:=~ -{\frac 1{12}}\bar D^2 \left( -{\frac 14} D^2P ~+~ i \pa_{\un a}H^{\un a}\right)~~~.~~~~~~~~~~~~~~~~~  }
\ee
The last two equations are the equations of motion of $H_{\un a}$ and $P$ from 4D supergravity.  This suggests the existence of fields strengths which correspond to the 5D corrections to these equations of motion:
\bea
\label{gravfs2a}
{ G}_{\un a}^\prime&:=&-2 \pa_5^2 H_{\un a}+2\pa_5\left(\bar D_{\dot \a}\J_\a-D_\a \bar \J_{\dot \a}\right)-i\pa_{\un a}\left(T-\bar T\right)\\
\label{gravfs2b}
{ R}^\prime &:=&  \pa_5^2 P- \pa_5\left(D^\a \J_\a+\bar D_{\dot \a}\bar \J^{\dot \a}\right)-{\frac 14}\left(D^2T+\bar D^2\bar T\right)~~~.
\eea
We should also look for something that could be the equation of motion for $\J_\a$.  Taking a combination similar to the kinetic terms for a gravitino multiplet leads us to:
\be
\label{gravfs4}  \eqalign{
\l_\a&:=~ \left({\frac 12}D^2 \,+ \, 2\bar D^2\right)\J_\a ~-~ \left(D_\a \bar D_{\dot \a} 
\,+\, 2\bar D_{\dot\a} D_\a\right)\bar \J^{\dot \a} \cr
&~~~~~~+~ D_\a\pa_5 P~+~ 2 \, \pa_5 \bar D^{\dot \a}H_{\un a}~~~.  }
\ee
Similarly to ${ W}_{\a\b\g}$, we find the following three field strengths which can not be equations of motion because of their index structure and mass dimension:
\bea
\label{gravfs3a}
{ F}_{\a\b}&:=& \pa_{(\a}{}^{\dot \a}\left[ \bar D_{\dot \a} \J_{\b)}-D_{\b)}\bar \J_{\dot \a}-\pa_5 H_{\b)\dot \a}\right] \\
\label{gravfs3b}
{ F}_{\a\b}^\prime&:=&\bar D^{\dot \a} D_{(\a}\left[ \bar D_{\dot \a} \J_{\b)}-\pa_5 H_{\b)\dot \a}\right] \\
\label{gravfs3c}
{ F}_{\a\b}^{\prime \prime}&:=& i\pa_{(\a}{}^{\dot \a}\left[ \bar D_{\dot \a}\J_{\b)}+D_{\b)}\bar \J_{\dot \a}+2 \pa_5 H_{\b)\dot \a}\right]+{\frac 12} \pa_5 [D_{(\a},\bar D^{\dot \a}]H_{\b)\dot \a}+ \bar D^2 D_{(\a}\J_{\b)}~~~~~
\eea
These nine field strengths form a fundamental set in the sense that any other field strength can be written as some combination of supercovariant derivatives on them.  The question of finding the complete set, will be addressed in the last section of this paper.

Similar to Maxwell theory, these field strengths satisfy certain identities.  We found many and here we present them in three groups.  These groups are based on how many degrees of freedom the identity removes on-shell.  The first group we call the ``algebraic identities".  These identities relate field strengths algebraically and therefore remove the most degrees of freedom off-shell.

\noindent{\bf Algebraic Identities:}
\bea
\label{dim1b}
i{ F}_{\a\b}-{ F}_{\a\b}^\prime+{ F}_{\a\b}^{\prime\prime}=0\\
\label{dim1a}
4i { F}_{\a\b}+2{ F}_{\a\b}^\prime-D_{(\a}\l_{\b)}=0
\eea
The second group of identities we call ``representation reducing".  These identities imply, with the use of the equations of motion, that certain field strengths become linear or chiral superfields on-shell.

\noindent{\bf Representation Reducing Identities:}
\bea
\label{dim1.5o}
\bar D_{\dot \a}{ R}=0\\
\label{dim1.5a}
\bar D^{\dot \a}{ G}_{\un a}-D_\a{ R}=0\\
\label{dim1.5b}
\bar D^{\dot \a}{ G}_{\un a}^\prime -D_\a { R}^\prime+\pa_5 \l_\a=0\\
\label{dim1.5c}
2 D^\b { F}^\prime_{\a\b}-\left( D^2 +\bar D^2\right) \l_\a- \left( D_\a \bar D_{\dot \a}+ 2\bar D_{\dot \a}D_\a\right)\bar \l^{\dot \a}= 0\\
\label{dim2b}
\bar D^2 { G}_{\un a}+4i\pa_{\un a}{ R}=0\\
\label{dim2d}
\bar D^2 { F}_{\a\b}+{\frac i4}\bar D^2 D_{(\a}\l_{\b)}=0\\
\label{dim2f}
-{\frac 14}\bar D^2 D^\a \l_\a+12c \pa_5 { R}=0\\
\label{dim3b}
-12c \pa_5^2{ R}+\bar D^2 \left({\frac 14}D^2 { R}^\prime+{\frac i2}\pa^{\un a}{ G}^\prime_{\un a}  \right)=0
\eea
Finally, we have the ``dynamical" identities.  After using the first two sets of identities, the dynamical identities and the equations of motion imply the correct equations of motion on the propagating field strengths.

\noindent{\bf Dynamical Identities:}
\bea
\label{dim2o}  
\bar D_{\dot \a} { W}_{\a\b\g}=~ 0 \\ 
\label{dim2a}
D^\a { W}_{\a\b\g} ~-~ i \,{\frac 12}\pa_{(\b}{}^{\dot \a}{ G}_{\g)\dot \a}=~0 \\
\label{dim2c}
\pa_{(\a}{}^{\dot \a}{ G}^\prime_{\b)\dot \a}~-~ 2 \pa_5 { F}_{\a\b}=~ 0\\
\label{dim2e}
\bar D^{\dot \a}D_{(\a}{ G}^\prime_{\b)\dot \a} ~-~ 2\pa_5{ F}^\prime_{\a\b} =~0
\\
\label{dim2g}
\pa^\b{}_{\dot \a}{ F}_{\a\b}~-~ \pa_\a{}^{\dot \b}\bar { F}_{\dot \a \dot \b} =~0 \\
\label{dim2.5}
-{\frac 14} \bar D^2 D_{(\a} { F}_{\b\g)}~+~ i \,24 \, \pa_5 { W}_{\a\b\g}=~0 \\
\label{dim3a}
\pa_5^2 { G}_{\un a} ~+~ \left[{\frac 14}\pa_5\left( {\frac 16}[D_\a, \bar D_{\dot \a}]
\,+\, i\pa_{\un a}\right)D^\b\l_\b +{\rm h.c.}\right]
~~~~~~~  \cr
+~ {\frac 1{16}}D^\b \bar D^2 D_\b { G}_{\un a}^\prime 
=~ 0 \\
\label{dim3x}
{\frac 18} D^\b \bar D_{\dot \a} { F}^\prime_{\a\b}~+~  \pa_5\left( {\frac 1{24}}  [D_\a, \bar D_{\dot \a}]\,+\, {\frac i4}\pa_{\un a}\right) D^\b \l_\b 
~~~~~~~ \cr
+ \pa_5 { G}_{\un a}~+~ {\frac 1{32}}D_\a \bar D^2 \bar \l_{\dot \a}
~+~ {\rm c.c.} =~ 0
\eea
Note that in Maxwell theory, we had no algebraic identities.  There were two representation reducing identities, (\ref{Maxbasic1}) and $\bar D_{\dot\a}W_\a=0$, and one dynamical identity (\ref{Maxbasic}).

As usual in supergravity, none of the field strengths have the correct mass dimension to appear quadratically in the Lagrangian that describes a Poincar\' e supergravity action.  We can almost construct a gauge invariant Lagrangian by contracting the field strengths with the prepotentials that have the same index structure.  Using this procedure, we arrive at the gauge invariant Lagrangian density given by \cite{Linch:2002wg}:
\bea
\label{gravlag0}
L=L_0+cL_1~~~~,
\eea
where
\bea
\label{gravlag1}
L_0&=&-{\frac 12}\int d^4\q \Big\{ H^{\un a} { G}_{\un a} +P({ R}+\bar { R})\Big\}~~~,
\eea
is simply the Lagrangian density for linearized old minimal supergravity \cite{Buchbinder:qv} and\footnote{It is possible to write the last line of $L_1$ in terms of a prepotential for $T$ and the field \newline $~~~~~\,$
strength ${ R}$, resulting in a contribution similar to the last term in (\ref{gravlag1}).}
\bea
\label{gravlag2}
L_1&=& {\frac 12}\int d^4\q \Big\{ H^{\un a} { G}^\prime_{\un a}+P { R}^\prime -( \J^\a \l_\a +\bar \J_{\dot \a} \bar \l^{\dot \a})\cr &~& \hspace{2in} - (T-\bar T) \Big[ (\S-\bar \S) -i \pa_{\un a} H^{\un a}\Big]\Big\}~~~.
\eea
As in Maxwell theory, there is a free parameter $c$ that can not be fixed using gauge invariance.  It was claimed in \cite{Linch:2002wg} that, after component projection and integration of auxiliary fields, the value $c=-{\frac 12}$ yields a 5D Lorentz invariant theory.  We will leave this coefficient arbitrary and show that standard superspace techniques reproduce this result with far greater efficiency.

The equations of motion in conjunction with the identities given above will imply that only the 5D Weyl tensor, curl of the gravitino and photon propagate.  The equations of motion resulting from the dynamical system (\ref{gravlag0}) are given by:
\bea
\label{graveom1a}
{{\d S}\over{\d H^{\un a}}}&=& -{ G}_{\un a}+c{ G}^\prime_{\un a} =0~~~,\\
\label{graveom1b}
{{\d S}\over{\d P}}&=&-\left({ R}+\bar { R}\right)+c{ R}^\prime=0~~~,\\
\label{graveom1c}
{{\d S}\over{\d \J^\a}}&=&\l_\a=0~~~,\\
\label{graveom1d}
{{\d S}\over{\d T}}&=&3c{ R}=0~~~.
\eea
The three equations of motion (\ref{graveom1b}-\ref{graveom1d}) and the algebraic identities imply that, on-shell, the only independent non-zero field strengths are $\{ { W}_{\a\b\g}, { G}_{\un a}, { F}_{\a\b}\}$\footnote{
${ F}_{\a\b}$ is the analog of the matter gravitino Weyl 
tensor denoted by ${ W}_{\a\b}$ in the work of reference \cite{Gates:1979gv}.
}.  We proceed to study the consequences of the representation reducing identities.  From (\ref{dim1.5a}) we see that ${ G}_{\un a}$ is transverse linear on-shell:
\bea
\label{Gred}
\bar D^{\dot\a}{ G}_{\un a}=0~~\Rightarrow~~
D^\a{ G}_{\un a}=D^2{ G}_{\un a}=\bar D^2{ G}_{\un a}
=\pa^{\un a}{ G}_{\un a}=0~~~.
\eea
The two identities (\ref{dim1.5c}) and (\ref{dim2d}) imply that $F_{\a\b}$ is also transverse linear on-shell:
\bea
\label{Fred}  
\bar D^2 F_{\a\b}=0~~,~~~~~~~~~~~~\cr
D^\a F_{\a\b}=0~~\Rightarrow~~D^2F_{\a\b} ~=~ 0 ~~.
\eea
Note that 4D massive superfields that satisfy (\ref{Gred}) or (\ref{Fred}) would have superspin-$\frac 32$ \cite{Massive2, Buchbinder:qv}.  We leave the investigation of the connection to massive 4D gauge theories to a future publication.  The remaining representation reducing identities are automatically satisfied on-shell and contain no further information.

After using all algebraic and representation reducing identities and three equations of motion, we are left with the chiral field strength, ${ W}_{\a\b\g}$, and the two transverse linear field strengths, ${ G}_{\un a}$ and ${ F}_{\a\b}$.  Using the equation of motion for $H_{\un a}$, (\ref{graveom1a}), and the dynamical identities, we will now show that this theory describes the correct propagating field strengths.  We begin by fixing the coefficient $c$.  To do this we look for an equation that produces the 5D d'Alembertian on one of these field strengths.  By differentiating (\ref{graveom1a}) and using (\ref{dim2a}), (\ref{dim2c}), and  (\ref{dim2.5}) we find:
\bea
\label{rep1}
{i\over 8\cdot 3!}\bar D^2 D_{(\a}\pa_{\b}{}^{\dot \b} (-{ G}_{\g)\dot \b}+c{ G}^{\prime}_{\g)\dot \b})=\left(\Box -2c\pa_5^2\right) { W}_{\a\b\g}=0~~~.
\eea
If ${ W}_{\a\b\g}$ is to be a non-trivial representation of the 5D Poincar\'e algebra, we must take $c=-{\frac 12}$.  Thus, the free parameter $c$ has been fixed without resorting to a component analysis.  We now turn our attention to finding the correct equations of motion for all propagating field strengths.

Starting with the lowest dimension field strengths which correspond to the vector gauge field, we look for the analog of the Maxwell equations (\ref{nosusyeom1}) and (\ref{nosusyeom2}) and the Bianchi identities (\ref{nosusybi1}) and (\ref{nosusybi2}).  The Bianchi identities are identical to the dynamical identities (\ref{dim2c}) and (\ref{dim2g}) if we set ${ F}_{\a\b}|=f_{\a\b}$ and ${ G}_{\un a}|=-\frac 12F_{5\un a}$.  With these definitions (\ref{nosusyeom2}) is satisfied by the representation reducing identities.  The equation of motion (\ref{nosusyeom1}) can be obtained by taking $-2\pa_5$ on (\ref{graveom1a}), using the dynamical identity (\ref{dim3x}) to substitute $F_{\a\b}^{\prime}$ terms for $\pa_5 { G}_{\un a}$, and then substituting the algebraic identity (\ref{dim1a}):
\be  \eqalign{
2\pa_5(\,{ G}_{\un a}\,+\, \frac 12{ G}_{\un a}^{\prime} \,)|
&=~ -\frac 1{16}D^\b\bar D_{\dot\a}F_{\a\b}^{\prime}|
~+~ \frac 1{16}\bar D^{\dot\b} D_\a \bar F_{\dot\a\dot\b}^{\prime}|
~+~ \pa_5 G_{\un a}^{\prime}|~~~\cr
&=~ \frac 12\pa_{\dot\a}{}^\b f_{\a\b}
~+~ \frac 12\pa_\a{}^{\dot\b} \bar f_{\dot\a\dot\b}
~+~ \pa_5 F_{5\un a}~~~.}
\ee
Thus, we have shown that the lowest components of ${ F}_{\a\b}$ and ${ G}_a$ satisfy the correct equations to be a spin-1 representation of the 5D Poincar\'e algebra.  These means that we have shown that both ${ F}_{\a\b}$ and ${ G}_a$ vanish under the 5D d'Alembertian.

The remaining on-shell field strengths ${ W}_{\a\b\g},~{ F}_{\a\b},$ and ${ G}_{\un a}$ all obey the 5D wave equation, thus, all of their components do as well. The remaining independent components are
\bea
\nonumber
C_{\a\b\g\d}:=D_{(\a}{ W}_{\b\g\d)}|~~,~~
C_{\a\b\g\dot \a}:={\frac 12}[D_{(\a},\bar D_{\dot \a}]{ F}_{\b\g)}|~~~,\\ 
\nonumber
C_{\a\b\dot\a\dot\b}:={\frac 12}[D_{(\b}, \bar D_{(\dot \b}]{ G}_{\a)\dot\a)}|~~~,~~~~~~~~~~~~~~~\\ 
\nonumber
f_{\a\b\g}^{(+)}:={ W}_{\a\b\g}|~~,~~
f_{\a\b\dot\a}^{(-)}:= D_{(\a}{ G}_{\b)\dot\a}| ~~~,~~~~~~~\\ 
\nonumber
{\bar f}^{(-)}_{\a\b\g}:=D_{(\a}{ F}_{\b\g)}|~~,~~
\bar f^{(+)}_{\a\b\dot \a}:= \bar D_{\dot \a} { F}_{\a\b}| 
~~~,~~~~~~~~~\\
\label{cfs2}
F_{\un a5}:={ G}_{\un a}|~~,~~
f_{\a\b}:={ F}_{\a\b}|~~~.~~~~~~~~~~~~~~
\eea
Since ${ F}_{\a\b}$ and ${ G}_{\un a}$ are transverse linear superfields, these are the only components we can obtain from these field strengths.  Further, $D^\a { W}_{\a\b\g}$ is proportional to $\pa_{(\a}{}^{\dot\a} { G}_{\b)\dot\a}$ and is not an independent propagating field strength.  Thus, the propagating field strengths are in one to one correspondence with (\ref{cfs1}),  on-shell.  

We would like to point out a hint about higher dimensional theories that is apparent from this discussion.  All higher dimensional extensions of minimal supergravity will share the 4D metric and gravitino structure that occurs in five dimensions.  The main difference will be the ``matter fields" which are necessary for supersymmetry in extra dimensions.  For example, in 6D, one requires a self dual 3-form field strength.  Therefore, we expect to see field strengths just like ${ G}_{\un a}$ and ${ F}_{\a\b}$ except that they will obey the correct equations to describe a 6D self dual 3-form.  In this way, the matter sectors will prove to be the best guide to constructing the gauge transformations of the prepotentials, and subsequently, the gauge invariant actions in higher dimensions.

This concludes our on-shell analysis.  In the final section we will discuss how to construct a closed and complete extension of these field strengths.
\section{Toward Extended Geometry}

$~~~\,~$To touch base with a full 5D super geometric description, let us recall the logic that we would use in such a construction.  First, a set of constraints are imposed on the algebra of covariant derivatives.  When constraints are set, the Bianchi identities imply that the torsions, curvatures and gauge field strengths are related by differential equations.  This means three things.  The first is that a set of tensors are set to zero identically.  Second, a set of tensors are now the higher components of a more fundamental set.  Finally, the tensors in the fundamental set are constrained in various ways.  These constraints are then solved by writing the fundamental super-field strengths as covariant derivatives acting on unconstrained prepotentials.  The prepotentials describe the fundamental gauge degrees of freedom and are used to construct actions.  In supersymmetric theories in dimensions higher than 4, the final step in this procedure, that of finding prepotential solutions to constraint equations, is not well understood.

	In the previous section, we examined the on-shell component field strength content of the 5D vector multiplet and minimal supergravity.  This analysis was based on a fundamental set of field strengths we will call $\cg$:  $\cg_{\rm VM}=\{{ W}_\a,~{ F}\}$ , $\cg_{\rm SG}=\{{ W}_{\a\b\g},~{ G}_a,~{ G}_a^{\prime},~{ F}_{\a\b},~{ F}_{\a\b}^{\prime},~{ F}_{\a\b}^{\prime\prime},~{ R},~{ R}^{\prime},~\l_\a\}$.  The higher components of these field strengths were related to each other by identities much like the usual Bianchi identities that would come from a complete geometrical formulation.  This structure mimics that of a constrained super geometry as discussed in the previous paragraph.  Thus, the  similarity of our identities to Bianchi identities means that there may be a way to connect this formalism to the dimensional reduction of a standard manifestly 5D supersymmetric formulation.  In this section, we wish to expand the fundamental set of field strengths to arrive at a set of field strengths that close under all combinations of covariant derivatives much like the usual Bianchi identities.  From the perspective of the previous section, this procedure will seem redundant.  We take the view point that our field strengths are a partial solution of the fully supersymmetric 5D Bianchi identities, and that our prepotentials are the full solution.  If this can be corroborated in the future, then we have found a way to solve higher dimensional supersymmetric constraint equations.

We will give a complete description of the method we use for extending the  geometry for super Maxwell theory.  This description includes a list of Bianchi identities which can be interpreted as a dimensional reduction of manifestly supersymmetric 5D super Maxwell theory.  The same description for supergravity is rather cumbersome and is not presented in this discussion.

Consider a fundamental set of field strengths $\cg$.  We want to complete this set.  By a complete set of field strength tensors we mean any set with the property that any tensor in the theory can be constructed from the fields in the set and their spacetime derivatives. To find a completion of $\cg$, we proceed by defining the derived set $\Hat \cg$ by taking fermionic covariant derivatives of the fields in $\cg$. In doing so, any time we find a field which cannot be written as an element of $\cg$ or its {\em spacetime} derivatives, we give it name and add it to $\Hat \cg$.  $\Hat \cg$ has a finite number of elements since the fermionic derivatives will eventually collapse into spacetime derivative or to zero. 

We now apply this method to $\cg_{\rm VM}$.  We begin the construction of the derived set by starting with the chiral spinor field strength.  There are two derived field strengths:
\bea
\label{Maxcompletion1}
{ f}_{\a\b}:=~ - i {\frac 12}D_{(\a}{ W}_{\b)}~~~,~~~ 
{ d}:=~ -{\frac 14} D^\a { W}_\a ~-~~ \frac 14\bar D_{\dot\a}\bar { W}^{\dot\a}~~~. ~~~
\eea
Here we note that because of the identity (\ref{Maxbasic1}), the derived field strength $ \bar \c_\a:=-{\frac 14}D^2 { W}_\a$ is the spacetime derivative of ${ W}_\a$.  The second fundamental field strength ${ F}$ is an unconstrained superfield.  This leads us to three derived field strengths:
\bea
\label{Maxcomplition2}
\j_\a:=~ D_\a { F}~~~,~~~\bar { M}:=~ -{\frac 14}D^2 { F}~~~,~~~ 
{ V}_{\un a}~=~{\frac 12}[D_\a ,\bar D_{\dot \a}]{ F}~~~.~~~
\eea
The identity (\ref{Maxbasic}) implies that the other two derived field strengths $\h_\a:=-{\frac 14}\bar D^2 D_\a { F}$ and ${ D}:={\frac 1{32}} \{ D^2, \bar D^2\} { F}$ are again spacetime derivatives of previously defined field strengths.

The full derived set is $\Hat \cg_{\rm VM} := \{{ f}_{\a\b}, { d}, \j_\a, \bar { M}, { V}_{\un a}\}$. To see that $\cg_{\rm VM}\cup \Hat \cg_{\rm VM}$ is a complete set of field strength tensors for 5D super-Maxwell, we will show that every combination of spinor derivatives acting on the union is a linear combination of spacetime derivatives of the union.  We believe that these identities are the dimensionally reduced Bianchi identities and we will present them using the standard organization based on mass dimension.  We also include in this list of identities any reality or chirality conditions, since these types of conditions come out of supergeometries quite frequently.

\noindent{\bf Dimension-1}
\bea
\label{maxdim1}
{ F} \,-\, \bar { F} ~=~ 0
\eea
\noindent {\bf Dimension-${\frac 32}$}
\bea
\label{maxdim1.5}
D_\a { F } ~=~  \j_\a
\eea
\noindent {\bf Dimension-2}
\bea
\label{maxdim2}
D_\a \j_\b ~=~ -2 \vep_{\a\b} \bar { M}~~~,~~~ \bar D_{\dot \a} \j_\a 
~=~ { V}_{\un a} ~-~ i \pa_{\un a} { F} ~~~,~~~{ V}_{\un a} ~-~ \bar { V}_{\un a}
~=~ 0~~~,~~~ \cr
~~~ D_\a { W}_\b ~=~  i { F}_{\a\b} ~-~ 2\vep_{\a\b} { d}~~~,~~~ \bar D_{\dot \a} { W}_\a
~=~ 0 ~~~,~~~
{ d}~-~ \bar { d} ~=~ 0 ~~~.~~~
\eea
\noindent{\bf Dimension-${\frac 52}$}
\bea
\label{maxdim2.5}
D_\a \bar { M} ~=~ 0~~~,~~~ D_{\a}  { M} ~=~  i\pa_{\un a}\bar \j^{\dot \a}
~-~ \pa_5 { W}_\a~~~,~~~
 \bar D_{\dot \b} { V}_{\un a}~ =~ 2 \vep_{\dot \b \dot \a}\pa_5 { W}_\a
~-~ i\pa_{\un a} \bar \j_{\dot \b} ~~~, ~~~\cr
 D_\g { F}_{\a\b}=2i \vep_{\g(\a} \pa_{\b)\dot \b} \bar { W}^{\dot \b} ~~~,~~~ \bar D^{\dot \g} { F}_{\a\b}~=~ -i 2\, \pa_{(\a}{}^{\dot \g} { W}_{\b)}~~~,~~~  D_\a { d}~=~ i {\frac 12}\pa_{\un a}\bar { W}^{\dot \a}~~~. ~~~
\eea
Thus, we have shown that $\cg_{\rm VM}\cup \Hat \cg_{\rm VM}$ is a closed set.  We would like to note in passing that most of the Bianchi identities (\ref{maxdim1}-\ref{maxdim2.5}) vanish when the theory is taken on-shell.  All of the non-trivial information is contained in $\cg_{\rm VM}$ and the two identities (\ref{Maxbasic1}) and (\ref{Maxbasic}).

We now turn to the linearized supergravity model.  Although we will not give the complete set of field strengths and Bianchi identities, we will discuss the complexity of the supergravity theory.  As before, we begin with the fundamental set $\cg_{\rm SG}$.  Let us contemplate the completion $\Hat \cg_{\rm SG}$. Since ${ W}_{\a\b\g}$ and ${ R}$ are chiral, they only contribute three fields each to $\Hat \cg_{\rm SG}$. However, ${ R}$, ${ G}$ and ${ G}^\prime$ are real, unconstrained and $\l$, ${ F}$, and ${ F}^\prime$ are complex, unconstrained superfields and they contribute $3\times 6$ and $3\times 9$ field strengths respectively.  This brings the number of (reducible) elements in $\Hat \cg$ to about 50.  This number will grow significantly upon performing the Lorentz decompositions.  The number of Bianchi identities is roughly 6 to 9 times the number of field strengths in the completed set.  This is of the correct order of magnitude to be equivalent to a dimensional reduction of a manifestly supersymmetric geometric description of 5D supergravity.  We would like to emphasize that the majority of the field strengths and Bianchi identities would vanish on-shell, leaving only the fundamental set and the related identities.  It is interesting to see the difference in magnitude between the number of off-shell and on-shell Bianchi identities.  This means that in higher dimensions we must be extremely careful when setting constraints,  since so many field strengths vanish on-shell, incorrectly constraining one field strength to zero may put the entire algebra on-shell.
\section{Conclusions}
$~~~\,~$We have presented the ${\cal N} = 1$ superspace geometry of 5D super-Maxwell theory and linearized supergravity.  The theories were taken on-shell and the correct propagating component field strengths were found.  Further, we have shown that it is possible to extend the fundamental set of field strengths to a complete set, which can be interpreted as the dimensional reduction of a manifestly 5D supersymmetric geometry.  In the future, we believe that explicit dimensional reduction will be shown to coincide with these geometries.

The most useful information that has been obtained from this analysis for supergravity, is the central role of the 5D vector field strength.  In any supergravity written in terms of ${\cal N} = 1$ superfields, the 4D graviton and gravitino will always be embedded in a chiral irreducible rank three spin-tensor ${ W}_{\a\b\g}$.  So, this part of the geometry is standard.  In higher dimensions, some other bosonic matter fields are required to complete the supersymmetry.  The matter field strengths will have the lowest mass dimension of all the propagating field strengths in the theory.  Thus, the superspace identities that the corresponding super-field strengths satisfy must take the same form as the higher dimensional component Bianchi identities.  This is exactly the case for 5D, where the component Bianchi identities (\ref{nosusybi1}) and (\ref{nosusybi2}) had the same form as the dynamical identities (\ref{dim2c}) and (\ref{dim2g}).  We can now make some non-trivial statements about higher dimensional supergravity theories.  For example, in 6D the matter field strength is a self-dual 3-form.  In 4D notation, this reduces to a vector field and a 2-form.  Although this is the same tensor structure of the 5D vector field strength, the Bianchi identities are completely different.  For the vector piece of the 6D 3-form field strength, one of the Bianchi identities is $\pa^{\un a}{ G}_{\un a}=0$. This is the equation defining an axial vector super-field strength. {\em This is an off-shell requirement.}  The only 4D supergravity that has this type of structure is new-minimal supergravity \cite{Gates:2003cz}.  Thus, we propose that 6D supergravity in ${\cal N} = 1$ superfields must look like new-minimal 4D supergravity coupled through $\pa_5$ and $\pa_6$ to matter.  One of those matter superfields is likely to be the gauge 3-form multiplet, but with the wrong sign for its kinectic energy.

One reason this rather unusual formulation of 4D, $\cal N$ = 1
supergravity is likely to emerge from a 3-brane in a 6D theory
is that such a construction could take advantage of a ``spin-0
to spin-0 Higgs mechanism" which allows the auxiliary 2-form of new
minimal supergravity to propagate.  This mechanism was shown to exist in \cite{GATNISH} when
new-minimal supergravity is coupled to chiral matter.  The
mechanism permits the auxiliary 2-form of new
minimal supergravity to ``eat'' a pseudo scalar field in the
3-form multiplet and thereby become physical. This is 
necessary in order that the physical axion  (presumably arising
from superstring/M-theory) should appear as a propagating state
on the three brane.

A new-minimal/3-form compensator supergravity theory would possess
some rather unique signatures;

(a.) It allows for a dilaton potential (albeit of a very restricted
form),

(b.) It possess a propagating axion in the form of a skew-symmetric
\newline $~~~~~~~~~~~$ 2-form, appropriate for implementation of the
Green-Schwarz \newline $~~~~~~~~~~~$ mechanism,

(c.) The auxiliary fields are $S$, $A_{\un a}$ and $C_{\un a \, \un
b \, \un c}$, where $C_{\un a \, \un b \, \un c}$ may be
\newline $~~~~~~~~~~~$ interpreted as the only remnant of 11D supergravity \newline $~~~~~~~~~~~$  (i.e.
M-theory),

(d.) In the absence of supersymmetry breaking there is a local
U(1) \newline $~~~~~~~~~\,~$ symmetry for which $A_{\un a}$ is
the gauge field.

Conceptually, the work in \cite{Linch:2002wg}
presents an interesting question.  Explicitly stated the question is
``How does the geometry of a higher dimensional superspace respond
to the presence of a brane?'' To our knowledge, this question has not
been answered previously.   All the preliminary evidence that
we have found in the present work, suggests that a complete set of 5D,
$\cal N$ = 1 superspace torsions and curvatures can likely be expressed
in terms of the non-linear extensions of the set of objects in ${\cal
G}_{\rm {SG}}$.  We expect that this type of reasoning will facilitate
the development of extensions of this theory to higher dimensions,
different backgrounds and to higher orders in fluctuations.

\section{Acknowledgments}
W.D.L. and J.P. would like to thank the organizers of TASI 2003 and the University of Colorado at Boulder for their hospitality and the stimulating environment wherein this project was conceived.  J.P. would like to thank Pierre Ramond and the Institute of Fundamental Physics at the University of Florida, where this paper was completed, for their warm hospitality.
\\[.5in]
{\it ``Hm...I'm not sure we wanna pay for a dimension we're not gonna use."}
\begin{flushright}
\href{http://www.gotfuturama.com/Multimedia/EpisodeSounds/1ACV03/18.mp3}{Fry, Futurama}
\end{flushright}


 \end{document}